\documentclass[sn-apa]{sn-jnl}

\usepackage{graphicx}%
\usepackage{multirow}%
\usepackage{amsmath,amssymb,amsfonts}%
\usepackage{amsthm}%
\usepackage{mathrsfs}%
\usepackage[title]{appendix}%
\usepackage{xcolor}%
\usepackage{textcomp}%
\usepackage{manyfoot}%
\usepackage{booktabs}%
\usepackage{algorithm}%
\usepackage{algorithmicx}%
\usepackage{algpseudocode}%
\usepackage{listings}%
\usepackage{etoolbox}
\usepackage{comment}

\theoremstyle{thmstyleone}%
\theoremstyle{thmstyletwo}%

\theoremstyle{thmstylethree}%

\begin{document}

\title[GPT-4V]{

Realizing Visual Question Answering for Education: GPT-4V as a Multimodal AI}



\author[1]{\fnm{Gyeong-Geon} \sur{Lee}}\email{ggleeinga@uga.edu}
\author*[1]{\fnm{Xiaoming} \sur{Zhai}}\email{xiaoming.zhai@uga.edu}









\affil*[1]{\orgdiv{AI4STEM Education Center}, \orgname{University of Georgia}, \orgaddress{\city{Athens}, \state{GA}, \country{United States}}}







\pretocmd{\abstractname}{}{}{}
\abstract{
Educational scholars have analyzed various image data acquired from teaching and learning situations, such as photos that shows classroom dynamics, students' drawings with regard to the learning content, textbook illustrations, etc. Unquestioningly, most qualitative analysis of and explanation on image data have been conducted by human researchers, without machine-based automation. It was partially because most image processing artificial intelligence models were not accessible to general educational scholars or explainable due to their complex deep neural network architecture. However, the recent development of Visual Question Answering (VQA) techniques is accomplishing usable visual language models, which receive from the user a question about the given image and returns an answer, both in natural language. Particularly, GPT-4V released by OpenAI, has wide opened the state-of-the-art visual langauge model service so that VQA could be used for a variety of purposes. However, VQA and GPT-4V have not yet been applied to educational studies much. In this position paper, we suggest that GPT-4V contributes to realizing VQA for education. By 'realizing' VQA, we denote two meanings: (1) GPT-4V realizes the utilization of VQA techniques by any educational scholars without technical/accessibility barrier, and (2) GPT-4V makes educational scholars realize the usefulness of VQA to educational research. Given these, this paper aims to introduce VQA for educational studies so that it provides a milestone for educational research methodology. In this paper, chapter II reviews the development of VQA techniques, which primes with the release of GPT-4V. Chapter III reviews the use of image analysis in educational studies. Chapter IV demonstrates how GPT-4V can be used for each research usage reviewed in Chapter III, with operating prompts provided. Finally, chapter V discusses the future implications.}




\maketitle
\section{Introduction}\label{introduction}

Real classrooms produce multimodal data - teachers and students not only speak, listen, read, and write but also behave, show, and see \cite{lee2023multimodality}. These multimodal data provide valuable insights into the dynamics of learning and teaching in classroom settings. Particularly, scholars have noted that the dynamics happening in the classroom, students' understandings on the learning content, and the represented image of certain people/objects could be fully revealed with the help of image data, such as classroom photos (which could be extended to videos), student-drawings, and textbook illustrations. Hence, educational scholars have long been analyzing and interpreting what the images imply for their focus of studies. It is notable that this so-called qualitative analysis of image data has been unquestioningly dependent on human researchers, with few attempts to automate it. Although not a few tasks related to educational studies, such as automatic scoring \citep{lee2024applying}, data augmentation \citep{fang2023using}, and question generation \citep{circi2023automatic}, are being automated using artificial intelligence (AI), most of these have focused on processing texts rather than images.

One of the reasons for this limited application of image processing methods in educational studies is the accessibility issue of the technology. Modern computer vision and image processing techniques rely on sophisticated machine learning architectures based on convolutional neural networks, which require programming skills to some extent. Nevertheless, educational scholars have limited access to them. Differently speaking, in turn, it implies that if the accessibility issue to the concurrent image processing technologies could be resolved, educational scholars will benefit much from those and further the innovation of the research methods toward ones that deal much more with multimodal data rather than just textual data.

Visual question answering (VQA), a relatively new technique in computer vision, enables users to ask a machine about a given image and receive the answer from it, both in natural language. Due to its revolutionary approach that allows a user without any technical knowledge about image processing to extract information from it, VQA has been forecast to be one of the most promising AI frameworks \citep{antol2015vqa}. The technical difficulty VQA has had until now was its limited capability of understanding and generating long and sophisticated natural language inputs and outputs. However, recently, large language models (LLMs) have been attached to computer vision models, constructing visual language models (VLMs) and significantly improving the potential of VQA techniques for various applications \citep{openai2023chatgpt}. Therefore, the only remaining barrier to the wide use of VQA techniques was the existence of the prominent VLM service open to the general public, which affords various applications from everyday to professional scalability.

GPT-4V (Generative Pre-trained Transformer 4-Vision), released by OpenAI in September 2023 as an extension module of ChatGPT, has opened the era of universal visual language model service, dramatically enhancing global users' experiences on VQA \citep{openai2023chatgpt}. After Microsoft Teams' initial reports on the performance and applicable domains of VQA with GPT-4V \citep{yang2023dawn}, scholars from medicine promptly tried GPT-4V and reported its performance in diagnosing symptoms and recognizing biological organs, mostly in a qualitative sense \citep{wu2023gpt4v, wang2023bioinformatics}. However, there have been very few reports of applying VQA for educational studies. As reviewed later, preemptive studies explored the possibility of utilizing VQA for educational studies with limited scope and empirical evidence.



In this paper, we explore how GPT-4V could contribute to realizing VQA for education. By 'realizing' VQA, we denote two meanings: (1) GPT-4V realizes the utilization of VQA techniques by any educational scholars without technical/accessibility barriers, and (2) GPT-4V makes educational scholars realize the usefulness of VQA to educational research. Given these, this paper aims to first introduce VQA for educational studies so that it provides a milestone for educational research methodology. Secondly, we exemplified five VQA tasks with GPT-4V, which pertains to instructional material, pedagogical practice, learner engagement, visual assessment, and classroom environment.



\section{Visual Question Answering: Development and Current Status}

In this section, we review VQA techniques, the characteristics of GPT-4V, and VQA used for educational studies.

    \subsection{Visual Question Answering} \label{Gyeong-Geon}

VQA is a relatively new task challenged by the AI initiative. \cite{antol2015vqa} proposed VQA as the new task for the machine learning models. They defined VQA as "given an image and a natural language question about the image, the task is to provide an accurate natural language answer" (p. 2425). \cite{antol2015vqa} framed VQA task as lying in the intersection of computer vision, natural language processing, and knowledge representation and reasoning. They also suggested that the new AI-required task needs to require multimodal knowledge and quantitative evaluation metrics in response to open-ended, free-form questions as well as multiple-choice questions. It was anticipated that to realize VQA, the machine needs to be equipped with fine-grained recognition, object detection, activity recognition, knowledge-based reasoning, and commonsense reasoning. For example, \cite{antol2015vqa}'s first VQA model could answer questions such as "Does this man have children?," "what is the woman reaching for?," and "Are the kids in the room the grandchildren of the adults?" for each of given image as like "yes," "glass," and "probably."

\cite{antol2015vqa}'s contribution to the VQA includes the construction of the benchmark dataset and the suggestion of promising machine learning architecture. Their VQA dataset compiled real images, abstract scenes, splits, captions, questions, and answers mostly from the Microsoft Common Objects in Context (MS COCO) dataset. The final VQA benchmark dataset included more than 250,000 images, more than 750,000 questions, and nearly 10,000,000 answers. \cite{antol2015vqa} trained their model with VGGNet and LSTM (Long Short-Term Memory) modules and reported their best model (LSTM Question + Image) showed a test accuracy of 54.06\%. Later studies have engaged in developing algorithms that show improved performance and/or suggesting other benchmark datasets to extend the applicability of VQA for various domains, such as medical studies (see \cite{lee2023multimodality} for a comprehensive review). However, there have been two main barriers to using VQA for educational studies: developing VQA models and specializing their functions for educational purposes demanded access to advanced machine learning technologies.

\subsection{GPT-4V}

\subsubsection{Visual Language Model}

Prior approaches to accomplishing VQA machine were successful to some extent but had weaknesses in generating informative and long-enough descriptions on the given image, due to their limited natural language processing, understanding, and generating capabilities. Therefore, efforts to significantly improve the machines' language generative performance were made. VLMs not only aim to understand and reason about visual and textual information, but also to generate sophisticated textual answers to the user's question. Briefly speaking, VLMs combine a computer vision model with a pre-trained LLM so that they tackle the barrier between visual and textual modalities \citep{zhang2024vision}. For example, in contrastive learning algorithms, which have been typically used to train milestone VLMs such as CLIP \citep{radford2021learning} and ALIGN \citep{jia2021scaling}, textual and visual data pairs are converted through encoders and mapped to each other. Meanwhile, VLMs such as VisualBERT with masked-language modeling \citep{li2019visualbert} and Flamingo with cross-attention mechanism \citep{Alayrac2022Flamingo} enabled generating more authentic natural language answers to the user's query. For example, if a user gives Flamingo an image of a dog and an image of a cat, saying "[Cute dog's image] This is a very cute dog. [Serious cat's image] This is," the machine returns "a very serious cat" \citep{Alayrac2022Flamingo}.

\subsubsection{GPT-4V and Its Features}

GPT-4 had originally been released as a powerful LLM, which supports textual modality \citep{openai2023gpt4}. However, in September 25th, 2023, \cite{openai2023chatgpt}  officially announced that "ChatGPT can now see" with the GPT-4V model, adding visual modality to become a multimodal AI model. In the technical sense, VisualChatGPT, a predecessor of GPT-4V from Microsoft, is known to have tried combining several visual foundation models such as Stable Diffusion as image processing module and GPT as text processing module \citep{wu2023visual}. Meanwhile, GPT-4V system card \citep{OpenAI2023GPT4V} stated, "incorporating additional modalities (such as image inputs) into large language models (LLMs) is viewed ... as a key frontier in artificial intelligence research and development" (p. 1). 

Right after the release of GPT-4V, Microsoft researchers published 'preliminary explorations with GPT-4V' \citep{yang2023dawn}. In the 166-page long comprehensive report, \cite{yang2023dawn} summarized GPT-4V's input modes, working modes and prompting techniques, vision-language capability, visual referring prompting as a method to interact with humans, etc. They examined  GPT-4V's capablility of as emerging VQA tasks, rather than quantitatively evaluating GPT-4V's performance based on existing benchmark datasets.

According to \cite{yang2023dawn}, first, GPT-4V inherits GPT-4 (no vision)'s unimodal text language capabilities. However, GPT-4V's genuine power is revealed when addressing multimodal image-text pair inputs to generated textual outputs - such as image recognition, object localization, image captioning, VQA, visual dialogue, dense caption, etc., even with multiple images combined with text prompt \citep{yang2023dawn}. They particularly appreciated GPT-4V's ability to address versatile tasks, some of which are even fabricated by the researchers rather than being benchmark training data, showing that GPT-4V has some generic visual reasoning abilities. According to \cite{yang2023dawn}, this expanded model and data scales differentiate GPT-4V from previous VLMs. They also noted that GPT-4V is uniquely capable of understanding visual markers on the input images, which suggests future development of genuine human-computer interaction formats such as visual referring prompting. Based on these results, \cite{yang2023dawn} foresaw that GPT-4V will be the "dawn" of multimodal foundational LLMs that will serve solving real-world problems.

Interestingly, there are some educational tasks among dozens of VQA examples from \cite{yang2023dawn}. For example, they shown that GPT-4V is able to understand tangram activity, read academic research paper, answer to the question related to the Earth's interior structure. Further, GPT-4V can solve several types of problems from the Wechsler Adult Intelligence Scale (WAIS), based on its visual reasoning ability.

Meanwhile, \cite{wu2023early} provided human evaluation results of GPT-4V as well as other VLMs, in terms of five tasks - visual understanding, language understanding, visual + language understanding, visual puzzle solving, and understanding of other modalities. While \cite{wu2023early} stressed that current automatic metrics are not suitable to evaluate GPT-4V's VQA performance, they stated that "GPT-4V performs really well on various tasks" when judged by human evaluators, outperforming some competing models such as Qwen-VL-Chat. However, they also revealed that the early version of GPT-4V shows difficulty in solving math problems.

Notably, \cite{OpenAI2023GPT4V} specified that GPT-4V was trained with concern for the safety properties, including harmful content, privacy, cybersecurity, etc. OpenAI developers and the red team achieved more than 90\% of the intended behaviors of GPT-4V in terms of disallowed behavior, person identification, ungrounded inference, and image jailbreak (e.g., CAPTCHA image decryption). However, it also turned out that GPT-4V sometimes provides scientifically or medically inappropriate information, which requires prudent use of it in terms of fact-checking. This shows the promise and possible pitfalls in utilizing GPT-4V for education.

GPT-4V has supported API (Application Programming Interface) since November 2023. Users with certain programming skills could call GPT-4V API for their purposes. However, even if users are unable to use API, they can still use GPT-4V's VQA functions on the ChatGPT chatbox interface with the subscription for GPT-4. This easy and user-friendly interface is very impactful for the universal use of GPT-4V.



\subsection{VQA Used for Educational Studies}

\subsubsection{Studies Before GPT-4V}

Originally, \cite{bigham2010vizwiz} first proposed VQA as the problem to be solved by technologies. \cite{bigham2010vizwiz} implemented a system in which visually impaired learners take photos of everyday life situations, upload them to the server, and receive real-time natural language responses from human helpers. For example, if visually impaired learners take a photo of a shelf with many cans and ask a question like 'Which one is the can of corn?,' they will soon receive the answer on their mobile device, which is from the real people helping the visually impaired through internet communication. Responding to \cite{bigham2010vizwiz}, VQA has been utilizing technological approaches to generate answers rather than relying on human, since \cite{antol2015vqa}.

However, even though VQA was suggested to practically serve people from its origin, there has been limited number of studies that focus on VQA technology for education. Several of them noted that VQA techniques can be used for textbook illustration analysis. \cite{aishwarya2022stacked} combined VGG16 as a vision model and Bidirectional Encoder Representations from Transformers (BERT) as a language model to construct a VLM model that showed 67.15\% accuracy for simple VQA tasks - e.g., "Q: Which animal is shown in the image" and "A: Bullock." With equivalent examples, \cite{gupta2023eduv} suggested that students can have a visual conversation with the VQA system about what is being represented in textbook illustrations. Meanwhile, \cite{suresh2018gamification} explored the possibility that VQA can be used for gamification, which allows early learners (3-4 years old) to answer whether a random image is about 'animal' or 'sport' (Level 1), or which animal or sport (Level 2). Other studies also reported that VQA system can be used as a chatbot that supports student learning of English, Mathematics, and Chinese \cite{lin2023research}  or concepts about weather and atmosphere \cite{sophia2021edubot}. However, most of them focused on reporting technical development process rather than educational implications.

    \subsubsection{Studies After GPT-4V}

    Several educational studies have been reported with regard to VQA after the release of GPT-4V in October 2023. Representatively, \cite{lee2023nerif} used GPT-4V for automatic scoring of student-drawn models in response to science items. They improved GPT-4V's image scoring accuracy based on the novel NERIF (Notation-Enhanced Rubric Instruction for Few-shot Learning) method. In consequence, they reported GPT-4V could assess students' drawn-models with mean accuracy = .51 with SD = .037, given with the multinomial scoring rubric.

Meanwhile, \citep{lee2023multimodality} comprehensively discussed the significance of multimodality of AI for education, which is a necessary component of artificial general intelligence, mentioning GPT-4V as the state-of-the-art multimodal AI. However, these studies took only a small portion of VQA's potential into consideration and did not suggest a comprehensive framework of VQA for education. 

    Recently, \cite{bewersdorff2024taking} rightly focused on the multimodal nature of GPT-4V. They suggested that multimodal LLMs could be used for content creation, supporting and empowering learning, and assessment and feedback, with several example prompts for GPT-4V. While their work provides insights into the affordances of multimodal LLMs, they have focused on a couple of aspects of science education, rather than education in general, and on teaching and learning, rather than educational research.

    In summary, VQA techniques used for educational studies have been conceptual and suggestive before GPT-4V and increased in number and applicability after GPT-4V - however, their scope and impacts have not reached the educational research, which frequently analyzes image data collected from classrooms, textbooks, teachers, and learners.

\section{GPT-4V Realizing VQA for Education}

In this section, we present GPT-4V's affordance it provides educators with its VQA functionality. The VQA application areas related to educational studies explored in this study are:

\begin{itemize}
    \item Eliciting applications of intriguing image
    \item Analyzing instructor's pedagogical practice
    \item Analyzing learners' engagement
    \item Assessing learner-generated visual data
    \item  Assessing classroom resources and environments
\end{itemize}

In the below subsections, we present exemplary utilizations of GPT-4V's VQA capabilities for each area, with the corresponding input image, user prompt, and the VLM's response.

\subsection{Eliciting Applications of Intriguing Image}

\begin{figure}
    \centering
    \includegraphics[width=1\linewidth]{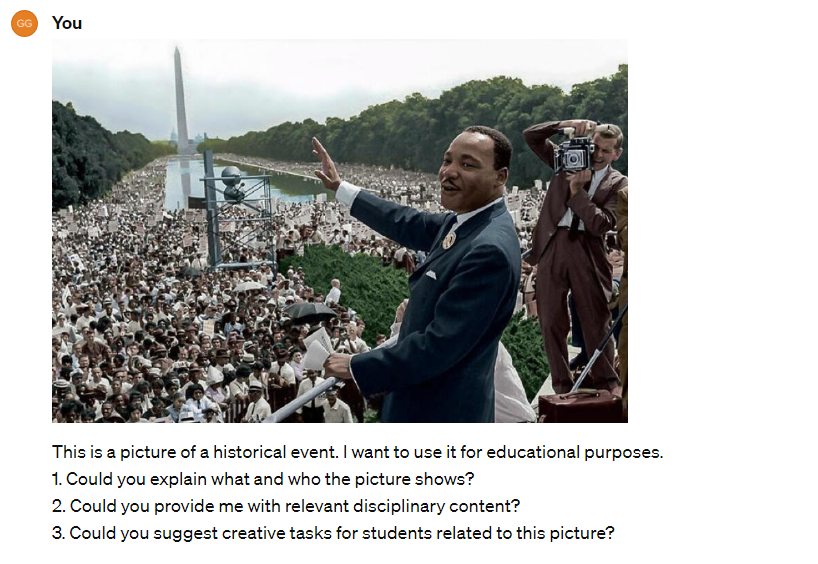}
    \caption{VQA Task 1: Eliciting Applications of Intriguing Image}
    \label{fig:mlk_address_question}
\end{figure}

\begin{figure}
    \centering
    \includegraphics[width=1\linewidth]{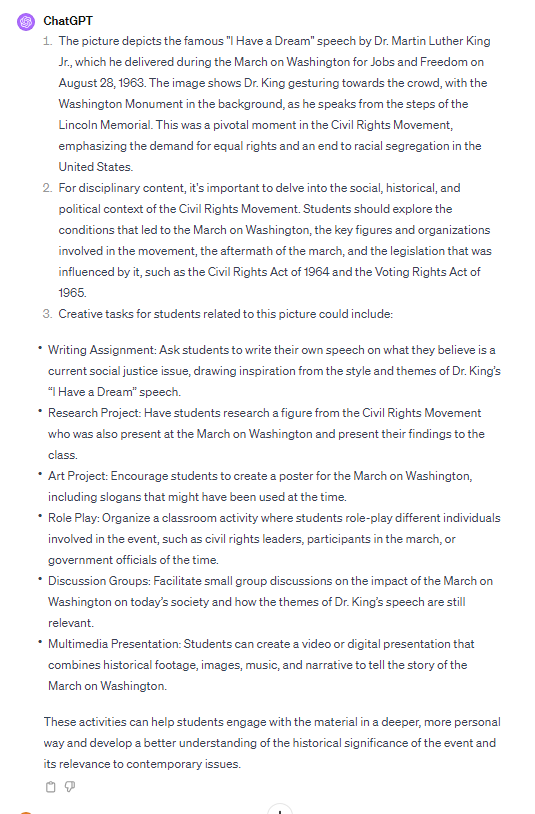}
    \caption{GPT-4V's Response to the VQA Task 1}
    \label{fig:mlk_address_answer}
\end{figure}

GPT-4V is not only able to properly understand the arbitrarily given image but also relate the picture with disciplinary content and pedagogy. The example of VQA for eliciting applications of intriguing image is presented as Figures \ref{fig:mlk_address_question}-\ref{fig:mlk_address_answer}. 

As the question (Figure \ref{fig:mlk_address_question}), the user gives a photo of Rev. Martin Luther King Jr.'s speech held in 1963, at the Lincoln Memorial in Washington, D.C. With the attached image, the user specifies that he/she wants to use it for educational purposes and asks GPT-4V to (1) explain what and who the picture shows, (2) provide relevant disciplinary content, and (3) suggest creative learning tasks related to it.

In response (Figure \ref{fig:mlk_address_answer}, GPT-4V (1) correctly says "the picture depicts the famous "I Have a Dream" speech by Dr. Martin Luther King Jr.," and provides precise information (date, place) and its historical background context (Civil Rights Movement; March for Washington on Freedom and Jobs). Strikingly, it rightly says it was "from the steps of the Lincoln Memorial," which is not visible in the given photo. (2) As disciplinary content, it recommends, "delve into the social, historical, and political context of the Civil Rights Movement." And "students should" study key figures, organizations, and legislation related to the march. (3) As creative learning tasks, it proposed writing assignments, research projects, art projects, role play, discussion groups, and multimedia presentations.

\subsection{Analyzing Instructor's Pedagogical Practice}

\begin{figure}
    \centering
    \includegraphics[width=1\linewidth]{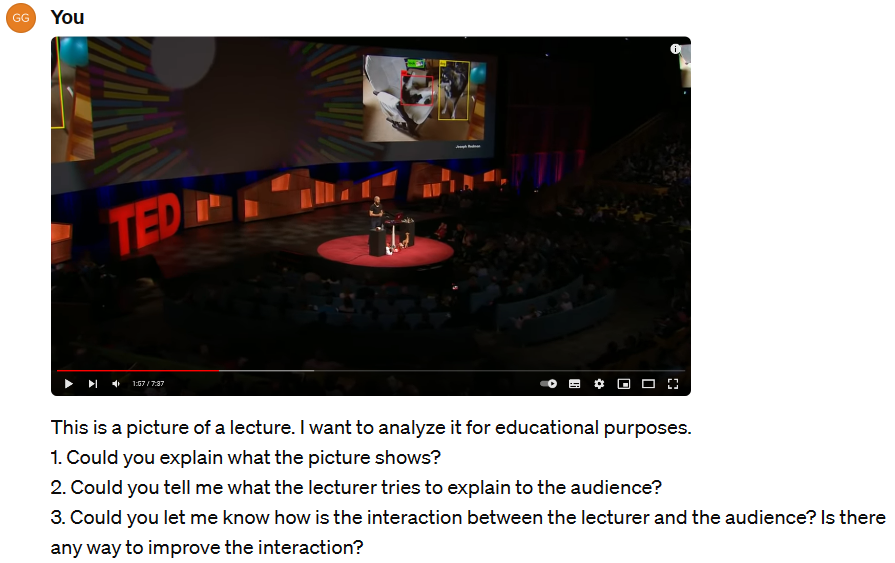}
    \caption{VQA Task 2: Analyzing Instructor's Pedagogical Practice}
    \label{fig:TED_question}
\end{figure}

\begin{figure}
    \centering
    \includegraphics[width=1\linewidth]{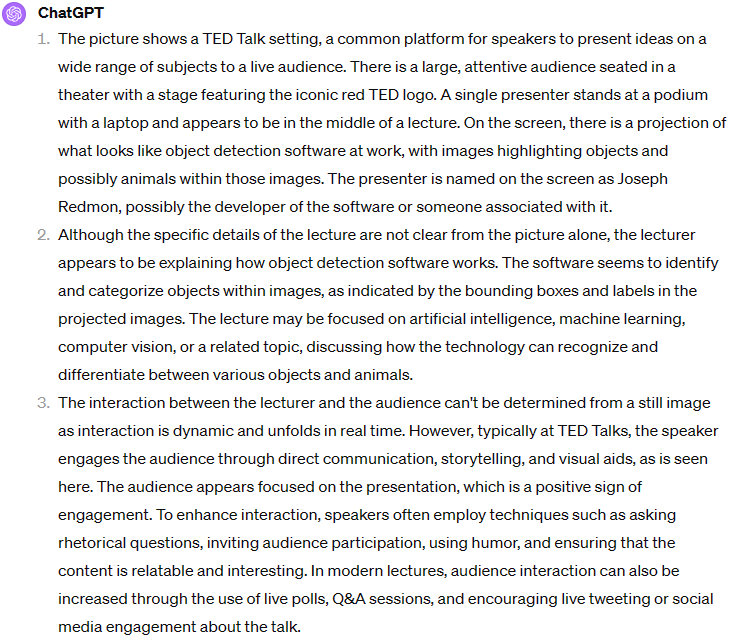}
    \caption{GPT-4V's Response to the VQA Task 2}
    \label{fig:TED_answer}
\end{figure}

GPT-4V understands where the teaching practice is being held, grasps which learning content is being delivered, describes the interaction between the instructor and learners, and suggests some ways to improve it. The example of VQA for analyzing instructor's pedagogical practices is presented as Figures \ref{fig:TED_question}-\ref{fig:TED_answer}.

As the question (Figure \ref{fig:TED_question}), the user gives a photo of a TED lecture by Joseph Redmon on YOLO (You Only Look Once) technique, which is a milestone real-time object detection method in computer vision, which is also relevant to VQA techniques. The picture was taken from its YouTube video.\footnote{https://www.youtube.com/watch?v=Cgxsv1riJhI; Accessed Mar 15, 2023} The image shows the lecturer on the podium, showing the share screen with a cat and a dog recognized by computer with red and yellow boxes, and audiences under the shadow around the spotlighted podium. With the attached image, the user specifies that he/she wants to use it for educational purposes and asks GPT-4V to (1) explain what the picture shows, (2) tell what the lecturer tries to explain, and (3) how is the interaction between the lecturer and the audience and the possible way to improve it.

As response (Figure \ref{fig:mlk_address_answer}, GPT-4V (1) saw that "the picture shows a TED Talk setting, a common platform for speakers to present ideas on a wide range of subjects to a live audience." and provides more information in the photo, as "A single presenter stands at a podium with a laptop and appears to be in the middle of a lecture," where "the presenter is named on the screen as Joseph Redmon." (2) Even with very little information given in the picture, GPT-4V correctly inferred that "the lecturer appears to be explaining how object detection software works," which is relevant to "artificial intelligence, machine learning, [or] computer vision," used for "identify and categorize objects within images, as indicated by the bounding boxes and labels in the projected images." (3) For instructor-learners interaction, GPT-4V described "typically at TED Talks, the speaker engages the audience through direct communication, storytelling, and visual aids, as is seen here," and "the audience appears focused on the presentation, which is a positive sign of engagement." Finally, it suggested several strategies to enhance interaction, such as "live polls, Q\&A sessions, and encouraging live tweet or social media engagement about the talk."


\subsection{Analyzing Learners' Engagement}

\begin{figure}
    \centering
    \includegraphics[width=1\linewidth]{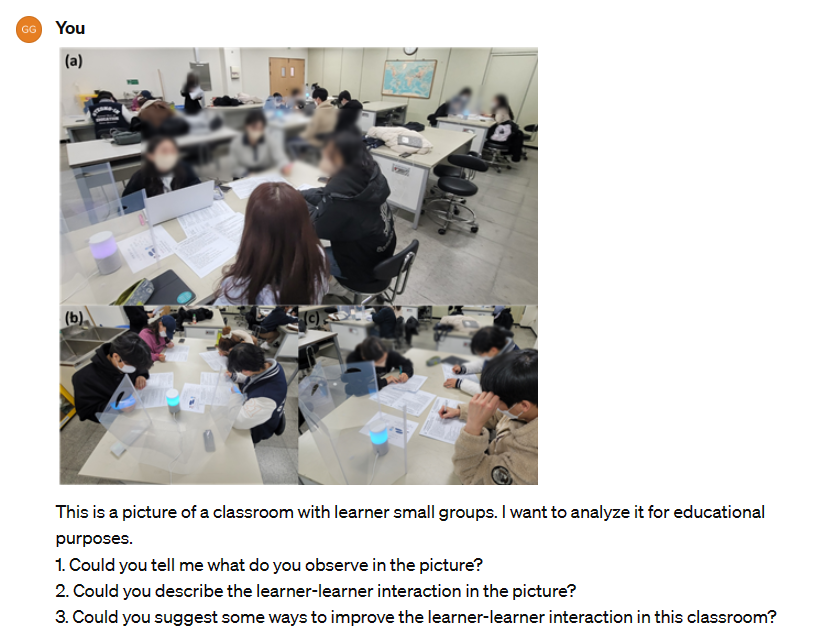}
    \caption{VQA Task 3: Analyzing Learners' Engagement (image adopted from \cite{Lee2024CLAIS})}
    \label{fig:inter_question}
\end{figure}

\begin{figure}
    \centering
    \includegraphics[width=1\linewidth]{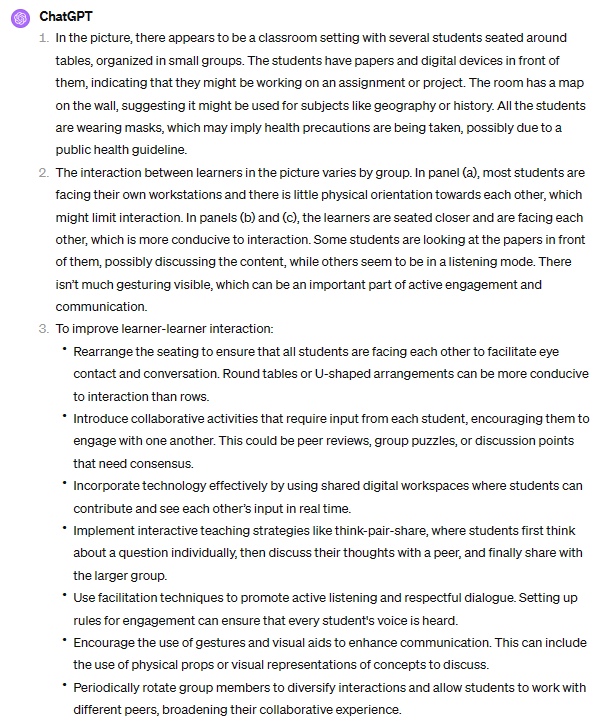}
    \caption{GPT-4V's Response to the VQA Task 3}
    \label{fig:inter_answer}
\end{figure}

GPT-4V detects the objects on the small group table, clearly figures out how learners' interaction is being formulated in the classroom, and suggest some ways to improve it. The example of VQA for analyzing learners' engagement is presented as Figures \ref{fig:inter_question}-\ref{fig:inter_answer}.

As the question (Figure \ref{fig:inter_question}), the user gives a photo of student small groups, going through a Jigsaw collaborative learning process with support of AI speakers (image adopted from \citep{Lee2024CLAIS}). The image shows students forming a small group with 4-5 members, one of which is AI speaker (with purple or blue lights in the image). With the attached image, the user specifies that he/she wants to analyze it for educational purposes, and asks GPT-4V to (1) explain what it observes, (2) describe the learner-learner interaction, and (3) suggest ways to improve the learner-learner interaction in the classroom.

As response (Figure \ref{fig:inter_answer}, GPT-4V (1) observed "a classroom setting with several students seated around tables, organized in small groups." And it correctly pointed out that "the students have papers and digital devices in front of them, indicating that they might be working on an assignment or project." It also mentioned that "all the students are wearing masks, ... possibly due to a public health guideline", which corresponds to the COVID-19 situation where the photo was taken. (2) GPT-4V specified that "In panel (a), most students are facing their own workstations and there is little physical orientation towards each other, which might limit interaction." In contrast, it stated that "in panels (b) and (c), the learners are seated closer and are facing each other, which is more conductive to interaction." Also, "some students are looking at the papers in front of them possibly discussing the content, while others seem to be in a listening mode."  This corresponds to the Jigsaw collaborative leaning process that was actually going on in Figure \ref{fig:inter_question} - (a) shows the overall classroom before the Jigsaw learning, (b) shows an expert group learning, and (c) shows a home group learning. However, it responded that "there isn't much gesturing visible, which can be an important part of active engagement and communication." Consequently, GPT-4V suggested to "encourage the use of gestures and visual aids to enhance communication," as well as to "rearrange the seating to ensure that all students are facing each other," "periodically  rotate group members to diversify interactions," etc.


\subsection{Assessing Learner-generated Visual Data}

\begin{figure}
    \centering
    \includegraphics[width=1\linewidth]{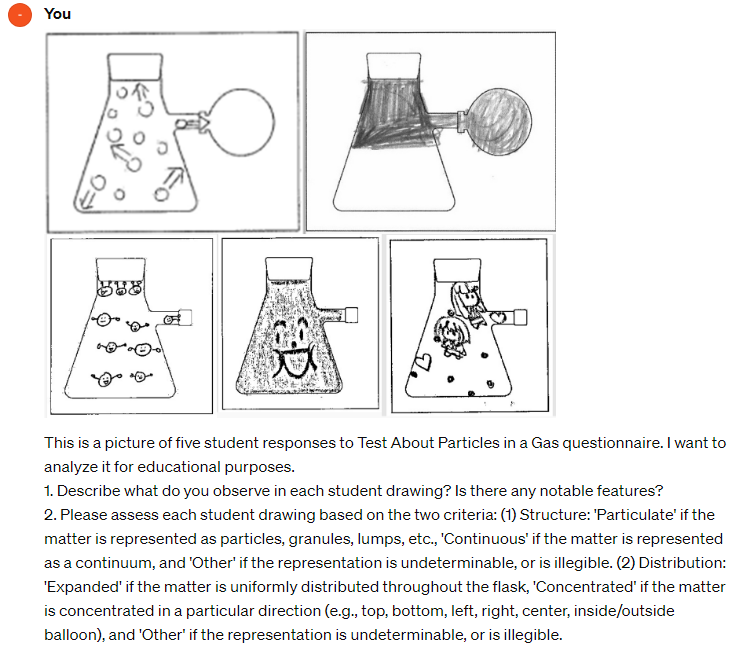}
    \caption{VQA Task 4: Assessing Learner-generated Visual Data (image adopted from \citep{lee2021particle}}
    \label{fig:drawing_question}
\end{figure}

\begin{figure}
    \centering
    \includegraphics[width=1\linewidth]{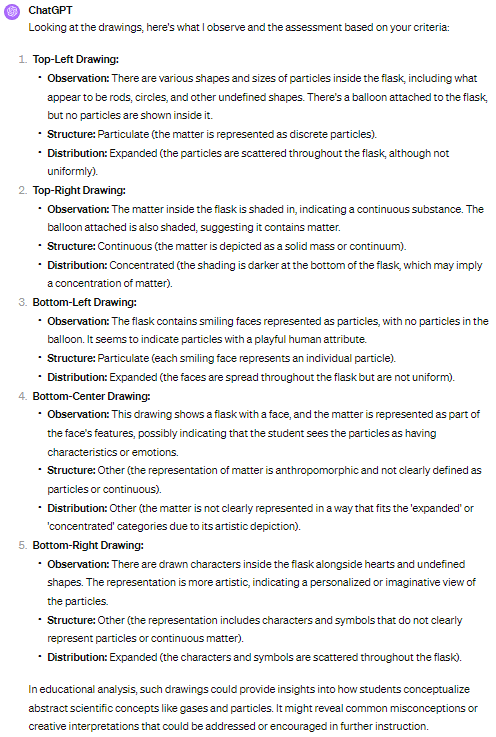}
    \caption{GPT-4V's Response to the VQA Task 4}
    \label{fig:drawing_answer}
\end{figure}

GPT-4V captures features in the student-generated visual responses, and assesses it according to the given scoring rubric with fair accuracy. The example of VQA for analyzing learner-generated visual data is presented as Figures \ref{fig:drawing_question}-\ref{fig:drawing_answer}.

As the question (Figure \ref{fig:drawing_question}), the user gives a five student response to Test About Particles in a Gas \citep{novick1981pupils}, drawing their mental models about how gas exist inside the flask in a pencil-and-paper format (image adopted from \citep{lee2023automated, lee2021particle}Lee et al., 2021). With the attached image, the user specifies that he/she wants to analyze it for educational purposes, and asks GPT-4V to (1) describe what it observes and whether there are notable features, and (2) assess each student drawing according to the two criteria: structure (particulate - continuous - other) and distribution (expanded - concentrated - other) which shows students' learning progression on the particulate nature of matter \citep{lee2023automated}.

As a response (Figure \ref{fig:drawing_answer}, GPT-4V provided the user with (1) what it observed for each student drawing - e.g., "there are various shapes and sizes of particles inside the flask" (top-left drawing), and "the matter inside the flask is shaded in, indicating a continuous substance" (top-right drawing). Notably, GPT-4V identified "smiling faces represented as particles" in the bottom-left drawing, "a flask with a face, and the matter is represented as part of the face's features, possibly indicating that the student sees the particles as characteristics or emotions" in bottom-center drawing, and "there are drawn characters inside the flask alonside hearts" in bottom-right drawing, each of which is particular to each drawing. (2) GPT-4V assessed each student drawing according to the given rubric with two criteria. GPT-4V assessed most aspect correctly - 60\% accurate for the structure aspect and 80\% accurate for distribution aspect, in comparison with human coders' decision. The unmatched cases were for the (a) structure aspect of bottom-center and bottom-right drawings, which human coders categorized both as 'expanded' - however, GPT-4V said that they were "anthromorphic " or "characters and symbols" which do "not clearly defined as particles or continuous." (b) Distribution aspect of bottom-center drawing, which human coders categorized as 'expanded' - however, GPT-4V said that "the matter is not clearly represented in a way that fits the 'expanded' or 'concentrated' categories due to its artistic depiction." It is also remarkable that GPT-4V said these drawings "might reveal common misconceptions or \textbf{creative interpretations} that could be addressed or encouraged in further instruction" (author emphasis).

\subsection{Assessing Classroom Resources and Environments}

\begin{figure}
    \centering
    \includegraphics[width=1\linewidth]{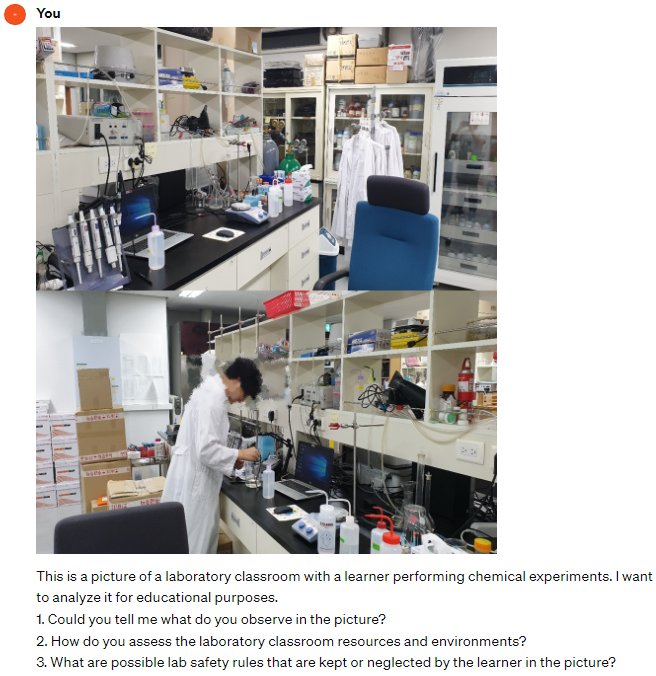}
    \caption{VQA Task 5: Assessing Classroom Resources and Environments}
    \label{fig:lab_question}
\end{figure}

\begin{figure}
    \centering
    \includegraphics[width=1\linewidth]{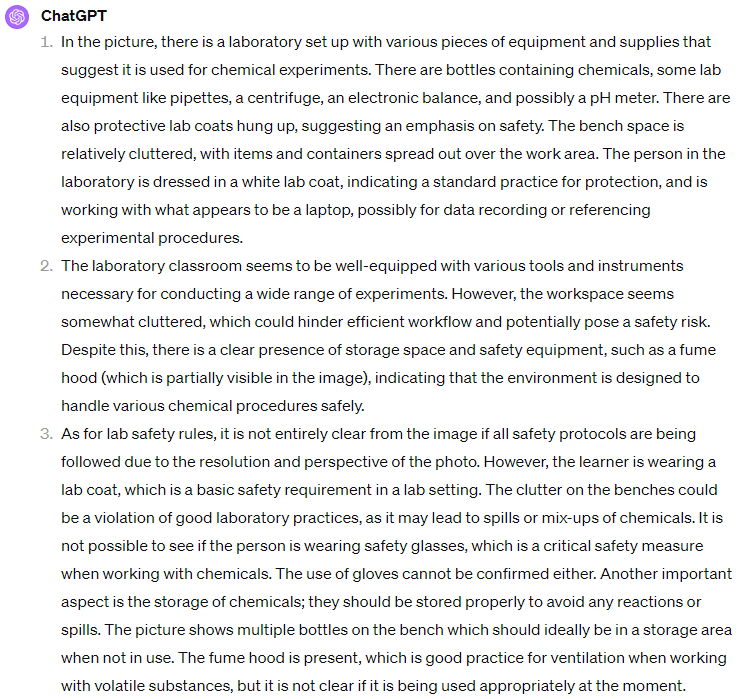}
    \caption{GPT-4V's Response to the VQA Task 5}
    \label{fig:lab_answer}
\end{figure}

GPT-4V detects the equipment and apparatus in the science laboratory classroom, assesses laboratory classroom resources and environments, and examines whether the lab safety rules are being secured or not. The example of VQA for assessing classroom resources and environment is presented as Figures \ref{fig:lab_question}-\ref{fig:lab_answer}.

As the question (Figure \ref{fig:lab_question}), the user gives a photo of laboratory classroom and a learner conducting chemical experiment in a laboratory. With the attached image, the user specifies that he/she wants to analyze it for educational purposes, and asks GPT-4V to (1) tell what it observes, (2) assess laboratory classroom resources and environments, and (3) what are possible lab safety rules that are kept or neglected in the picture.

As response (Figure \ref{fig:lab_answer}, GPT-4V (1) observed "a laboratory set up with various pieces of equipment and supplies," such as "bottles containing chemicals," "pipettes", "a centrifuge," "an electronic balance," and "possibly a pH meter." It also said "there are also protective lab coats hung up, suggesting an emphasis on safety" but "the bench space is relatively cluttered." "the person in the laboratory is dressed in a white lab coat," and there is "a laptop, possibly for data recodring or referencing experimental procedures." Its response was correct in general. (2) GPT-4V assessed the lab classroom resources and environment as "well-equipped with various tools and instruments necessary for conducting a wide range of experiment." However, it was noted that "the workspace seems somewhat cluttered, which could hinder efficient workflow and potentially pose a safety risk." Nevertheless, GPT-4V noticed "there is a clear presence of storage space and safety equipment, such as a fume hood," which indicates "the environment is designed to handle various chemical procedures safely." (3) For safety rules kept or neglected, GPT-4V rightly said "the learner is wearing a lab coat," "the clutter on the benches could be a violation of a good laboratory practices, as it may lead to spills or mix-ups of chemicals." GPT-4V said "it is not possible to see if the person is wearing safety glasses [since there was a mosaic on the learners' face], and "the use of gloves cannot be confirmed either" - actually, the learner was not wearing safety glasses or gloves. GPT-4V also mentioned about storage of chemicals, bottles on the bench, and fume hood, which needed to be appropriately stored or used.

\section{Discussions}

In step with the rapidly changing technological landscape around generative AI, this study examined the promises of VQA for educational researchers and practitioners, focusing on GPT-4V as a multimodal AI. Particularly, GPT-4V has been spotlighted as the state-of-the-art, a user-friendly multimodal AI service that is affordable.

Primarily, we tried to investigate how VQA techniques started, developed, and reached their current state as of 2024. Starting about 10 years ago \citep{antol2015vqa}, VQA techniques became widely available due to the rise of various VLMs. Especially, GPT-4V released in the fall of 2023 significantly impacted the availability of VQA for general users. As of April 2024, GPT-4V seems to be the state-of-the-art VLM, compared to other competing models. While there have been some studies that suggested VQA could be used for educational studies, recent studies have stressed the probable impacts of GPT-4V educational studies. The results of this review clearly show that there needs to be timely explorations of the full potential of GPT-4V, particularly for VQA, which could serve for education. This study can be the initiating attempt for the continuous efforts to integrate multimodal AI in education, which should follow in the near future. If this study has been somewhat scoping, qualitative, and initiating, further explorations of the benefits of VQA for education could be more comprehensive, quantitative, and fruitful.

Also, we selected five example image analysis tasks that could be used for educational studies, devised the input (image + prompt), and reported the GPT-4's response to the user queries that show its VQA capability.

The first  and second examples (eliciting applications of intriguing image, and analyzing instructor's pedagogical practice) have the potential to serve teachers. GPT-4V could help teachers plan and deliver lessons with intriguing pictures and a given topic through its correlating functions. GPT-4V also has the potential to provide feedback to instructors in the given specific context and situation. Since it is possible to transfer consecutive images from a video to GPT-4V, a glimpse of the real-time feedback to instructional practice using VQA is posited here, which needs future research.

The third and fourth examples (analyzing learners' engagement, and assessing learner-generated visual data) have the potential to serve learners. It is noteworthy that GPT-4V could grasp the aspect of learners' engagement through their physical orientation (including where they are facing) and gestures. Therefore, GPT-4V could help learners reshape their peer interactions within small groups by analyzing their configuration. The report of GPT-4V's automatic scoring ability on student-drawn models echoes previous studies (\citep{lee2023nerif, lee2023automated}. However, this study further reports novel features of GPT-4V's automatic scoring pattern -  which even apprehends students' "misconception" around the anthropomorphism of gas substances. Particularly, GPT-4V"s capability of acknowledging student drawings' "creative" points should be spotlighted. This provides empirical objection to some comments on AI's incapability of assessing students' creativity \citep{li2023can}. It is worth being mentioned again that GPT-4V even suggested that students' "creative interpretations ... could be ... encouraged in further instruction." Therefore, the result implies that GPT-4V and future VLM models could be used for automatic scoring not only aiming accurate evaluation according to rubric, but also for more qualitative and in-depth analysis of student-drawn answers.

The fifth task (assessing classroom resources and environments) has the potential to serve overall classroom. GPT-4V could detect not only frontal objects on the desk, it could figure out there is a fume hood, which only a small part of it was visible in the picture, showcasing GPT-4V's visual reasoning ability. Whereas in this study GPT-4V could detect various laboratory objects and assess the classroom environment based on those, this could be easily extended to usual classroom environments which have much fewer peculiar objects but more general objects such as "map on the wall, suggesting it might be used for subjects like geography or history" (see Figure \ref{fig:inter_answer}).

These results exhibit the exemplary types of VQA tasks that GPT-4V is capable of and its performance, which could be used for educational purposes in both research and practice. This study seems to be one of the first reports of collective VQA tasks serving educational purposes. Particularly, it is worth emphasizing that this study showed that a state-of-the-art VLM—i.e., GPT-4V—can perform those VQA tasks without any model training or fine-tuning from educational researchers/practitioners. Of course, while this study focused on examining the performance of GPT-4V on educational VQA, other VLM models can also be used for the same purpose. For example, open-source VLMs such as BLIP and LLaVa can be alternatives for GPT-4V according to the user's preference. Nevertheless, the utilization of VLM services that are highly accessible to broader users can dramatically change how teachers and/or researchers leverage the image data generated in the classroom, which was previously volatile and could not be fully exploited for understanding and improving classroom situations. Therefore, by facilitating visual data analysis, in addition to comparatively easier textual data analysis, VQA will consequently contribute to multimodal instruction.

\section {Conclusion}\label{}

In this study, we introduced the VQA techniques that could be utilized for educational purposes, reviewed its development and current status, pointed out image analysis in educational studies, and exemplified how GPT-4V could be used to realize VQA for educational studies. The results of this study qualitatively show the promising performance of GPT-4V in providing VQA function, that could be used for analyzing images for pedagogy. The implications of this report pertain to the following aspects. 

First, in research, VQA functions exemplified in this study could be introduced to draft ideas for image analysis in educational studies. Also, if researchers are willing to consider GPT-4V's reasoning ability as significant, it could contribute to triangulating the image analysis, providing an alternative viewpoint on the image other than from researchers, thanks to the large corpus and image dataset used to train GPT-4V. 

Second, in teaching, practitioners could use VQA with simple prompts to draft their judgment on the current status of individual learners, small groups, and the classroom as a whole. Of course, the final assessment should be done by the human instructor, who could take responsibility to the instructional decisions. If appropriately used, VQA technique could dramatically reduce the time of teachers needed to deal with multimodal data (e.g., in scoring constructed response questions) and will allow them to other tasks such as increased intervention time.

Finally, it should be emphasized again that this study is in step with the development of multimodal artificial general intelligence. While this study focused on VQA, which breaks the barrier between the two modes (textual and visual data), the AI initiative pursued by OpenAI and any other stakeholders is heading towards genuine multimodal intelligence that incorporates speaking and listening abilities \citep{lee2023multimodality}, and even further to embodied agent \citep{ortiz2024figure}. Therefore, VQA is the cornerstone for multimodal AI for education, but it is only the beginning of the end.

\section*{Acknowledgement}
This paper was funded by National Science Foundation(NSF) (Award no. 2101104, 2138854). Any opinions, findings, conclusions, or recommendations expressed in this material are those of the author(s) and do not necessarily reflect the views of the NSF.

\bibliography{sn-bibliography}

\end{document}